\documentclass[sigconf, screen]{acmart}

\usepackage[utf8]{inputenc}
\usepackage[T1]{fontenc} 

\usepackage{natbib}

\usepackage[framemethod=tikz]{mdframed}
\usetikzlibrary{shadows}
\newmdenv[tikzsetting={draw=gray,fill=white},
          roundcorner=5pt,shadow=true]{mdboxshad}
\mdfsetup{%
middlelinewidth=1.5pt
}

\usepackage{algorithmic}
\usepackage{algorithm}
\usepackage{array}
\usepackage[caption=false,font=normalsize,labelfont=sf,textfont=sf]{subfig}
\usepackage{textcomp}
\usepackage{stfloats}
\usepackage{amsmath}
\usepackage{url}
\usepackage{verbatim}
\usepackage{graphicx}

\makeatletter
\def\zifour@spc{\hyphenchar\font=`\-\relax}
\makeatother
\usepackage{hyperref}
\usepackage{csquotes}
\usepackage{enumitem}
\setlist[itemize,enumerate]{leftmargin=7mm}

\usepackage{colortbl}
\usepackage{xcolor}
\usepackage{multirow}
\usepackage{tabularx}
\usepackage{booktabs}
\usepackage{comment}
\usepackage[super]{nth}
\usepackage{hyperref}
\usepackage{listings}

\definecolor{background-gray}{gray}{0.96}
\usepackage[framemethod=tikz]{mdframed}
\newmdenv[
  topline=false,
  bottomline=false,
  rightline=false,
  skipabove=\baselineskip,
  skipbelow=\baselineskip,
  innertopmargin=6pt,
  innerbottommargin=6pt,
  innerleftmargin=12pt,
  innerrightmargin=12pt,
  tikzsetting={draw=black, line width=4pt},
  linecolor=black,
  backgroundcolor=background-gray
]{results}

\setcopyright{cc}
\setcctype{by}
\acmDOI{10.1145/3805760.3814922}
\acmYear{2026}
\copyrightyear{2026}
\acmISBN{979-8-4007-2601-9/2026/07}
\acmConference[AIware '26]{Proceedings of the 3rd ACM International Conference on AI-Powered Software}{July 6--7, 2026}{Montreal, QC, Canada}
\acmBooktitle{Proceedings of the 3rd ACM International Conference on AI-Powered Software (AIware '26), July 6--7, 2026, Montreal, QC, Canada}
\acmSubmissionID{fseaiware26main-pp028-data-p}
\received{2026-02-15}
\received[accepted]{2026-03-28}

\begin{document}

\title{A Dataset of Agentic AI Coding Tool Configurations}

\author{Matthias Galster}
\orcid{0000-0003-3491-1833}
\affiliation{%
  \institution{University of Bamberg}
  \city{Bamberg}
  \country{Germany}
}
\email{mgalster@ieee.org}

\author{Seyedmoein Mohsenimofidi}
\orcid{0009-0009-1620-2735}
\affiliation{%
  \institution{Heidelberg University}
  \city{Heidelberg}
  \country{Germany}
}
\email{s.mohsenimofidi@uni-heidelberg.de}

\author{Levi Böhme}
\orcid{0009-0000-7962-0111}
\affiliation{%
  \institution{University of Bayreuth}
  \city{Bayreuth}
  \country{Germany}
}
\email{levi.boehme@uni-bayreuth.de}

\author{Jai Lal Lulla}
\orcid{0009-0005-0024-8238}
\affiliation{%
  \institution{Singapore Management University}
  \city{Singapore}
  \country{Singapore}
}
\email{jailal.l.2025@phdcs.smu.edu.sg}

\author{Muhammad Auwal Abubakar}
\orcid{0009-0006-1028-0650}
\affiliation{%
  \institution{University of Bamberg}
  \city{Bamberg}
  \country{Germany}
}
\email{muhammad.abubakar@uni-bamberg.de}

\author{Christoph Treude}
\orcid{0000-0002-6919-2149}
\affiliation{%
  \institution{Singapore Management University}
  \city{Singapore}
  \country{Singapore}
}
\email{ctreude@smu.edu.sg}

\author{Sebastian Baltes}
\orcid{0000-0002-2442-7522}
\affiliation{%
  \institution{Heidelberg University}
  \city{Heidelberg}
  \country{Germany}
}
\email{sebastian.baltes@uni-heidelberg.de}

\renewcommand{\shortauthors}{Galster et al.}

\begin{abstract}
Agentic AI coding tools such as Claude Code and OpenAI Codex execute multi-step coding tasks with limited human oversight. To steer these tools, developers create repository-level configuration artifacts (e.g., Markdown files) for configuration mechanisms such as \textsc{Context Files}, \textsc{Skills}, \textsc{Rules}, and \textsc{Hooks}. There is no curated dataset yet that captures these configurations at scale. This dataset, collected from open-source GitHub repositories, fills that gap. We selected 40,585 actively maintained repositories through metadata filtering, classified them using GPT-5.2 to identify 36,710 as belonging to engineered software projects, and systematically detected configuration artifacts in these repositories. The dataset covers 4,738 repositories across five tools (Claude Code, GitHub Copilot, OpenAI Codex, Cursor, Gemini) and eight configuration mechanisms. We collected 15,591 configuration artifacts, the full content of 18,167 configuration files associated with these configuration artifacts, and 148,519 AI-co-authored commits. The dataset and the construction pipeline are publicly available on Zenodo under CC~BY~4.0. An interactive website allows researchers to browse and explore the data. This data supports research on context engineering, AI tool adoption patterns, and human-AI collaboration.
\end{abstract}



\maketitle

\section{Introduction}
\label{sec:introduction}

Agentic AI coding tools have changed how developers write software~\cite{DBLP:journals/corr/abs-2509-06216}. Tools such as OpenAI Codex, Claude Code, and Cursor operate with greater autonomy than traditional AI coding assistants: they plan multi-step tasks, execute code, run tests, and iterate on results with minimal human intervention~\cite{DBLP:journals/corr/abs-2602-09185, DBLP:journals/tosem/DongJJL24}. These tools break down high-level goals into subtasks and adjust their approach based on real-time feedback~\cite{DBLP:journals/corr/abs-2505-10468}. For instance, when asked to implement a new feature, such a tool may identify the relevant parts of the codebase, apply the necessary modifications, run tests, and refine its output based on the results.

As agentic tools see growing adoption~\cite{DBLP:journals/corr/abs-2601-18341}, developers have started to configure their behavior through \emph{configuration mechanisms}, i.e., means by which developers tailor tool and agent behavior to a project or workflow. Prior work~\cite{galster2026configuring} identified and defined eight configuration mechanisms as follows: \textsc{Context Files} are Markdown files loaded into the context of each session; \textsc{Skills} bundle reusable knowledge and invocable workflows; \textsc{Subagents} are specialized agents that operate in parallel to the central agent loop, in their own context; \textsc{Commands} are user-triggered shortcuts for predefined prompts; \textsc{Rules} are system-level instructions to control agent behavior; \textsc{Settings} are JSON/TOML configurations for project-level tool behavior; \textsc{Hooks} are scripts executed at specific agent lifecycle points; and \textsc{MCP} configurations register external tool or data connections via the Model Context Protocol.

A \emph{configuration artifact} is a tangible instance of a mechanism~\cite{galster2026configuring}: either a single configuration file (e.g., \texttt{CLAUDE.\allowbreak{}md}, or one subagent file such as \texttt{.\allowbreak{}claude/\allowbreak{}agents/\allowbreak{}reviewer.\allowbreak{}md}) or a directory bundling several configuration files that together define one artifact (e.g., a skill whose directory contains \texttt{SKILL.\allowbreak{}md} alongside scripts, references, and assets). 
Through configuration artifacts, developers communicate project constraints, coding conventions, and architectural decisions to AI agents~\cite{Seyedmoein2026, DBLP:journals/corr/abs-2511-12884}. These artifacts are version-controlled and collaboratively maintained. They represent a new category of software engineering documentation written for AI agents rather than human teammates, functioning as specifications that encode constraints and expected behaviors to guide code generation and modification.

Despite this growing adoption, no curated dataset systematically captures how developers configure agentic AI coding tools across open-source repositories. Without such data, researchers cannot study questions such as which configuration mechanisms developers actually use, how configurations differ across tools and programming languages, or how AI-assisted contributions relate to project characteristics or impact project quality.

We address this gap with a dataset of agentic AI coding tool configurations collected from GitHub. This work builds on a previous study~\cite{galster2026configuring}, which identified a taxonomy of eight configuration mechanisms and characterized their adoption across 2,853~repositories. Here, we contribute the underlying data and construction pipeline as reusable research artifacts with expanded scope and new data types: (1)~a curated sampling frame of 40,585 actively maintained GitHub repositories with rich metadata; (2)~LLM-based classification of 36,710 repositories as belonging to \emph{engineered} software projects, including justifications for the classification; (3)~15,591 configuration artifacts from 4,738 repositories across five tools and eight configuration mechanisms, together with the full content of 18,167 configuration files; (4)~148,519 AI-co-authored commits from repositories that use agentic AI coding tools; and (5)~a fully reproducible construction pipeline and an interactive website for exploring the dataset.

\section{Related Work}
\label{sec:related-work}

Work related to this dataset can be categorized into studies of AI-assisted development behavior, analyses of configuration artifacts, and existing datasets on AI-generated code.

\citeauthor{DBLP:journals/corr/abs-2601-18341}~\cite{DBLP:journals/corr/abs-2601-18341} measured coding-agent adoption on GitHub through file-, commit-, and pull-request-level heuristics over 128,018 repositories, analyzing adoption rates and the size and type of AI-assisted commits. Their file categorization distinguishes structured \emph{configuration files} (such as YAML) from natural-language \emph{rules and guidance files} (including \texttt{CLAUDE.\allowbreak{}md}, \texttt{AGENTS.\allowbreak{}md}, and Cursor rules) and is coarser than our eight-mechanism taxonomy. Their sample applies activity and size thresholds (non-fork, 5,000 lines of code, 100 commits, recent activity) without an engineered-project classification step, so their population includes tutorial and educational repositories that our engineered-project filter excludes; the two populations are complementary and support comparative analysis. \citeauthor{DBLP:journals/corr/abs-2509-14745}~\cite{DBLP:journals/corr/abs-2509-14745} and \citeauthor{DBLP:journals/corr/abs-2511-04824}~\cite{DBLP:journals/corr/abs-2511-04824} studied AI-authored pull requests and agentic refactoring, respectively. Both examined what AI tools produce, not how developers configure them. \citeauthor{he2026speed}~\cite{he2026speed} analyzed Cursor's effect on development velocity and code complexity for a single tool without examining configuration artifacts. \citeauthor{jiang2026beyond}~\cite{jiang2026beyond} studied Cursor rule files, covering one configuration mechanism for one tool. These studies produced empirical findings but not reusable, multi-tool datasets.

Two studies target configuration artifacts directly. \citeauthor{DBLP:journals/corr/abs-2511-12884}~\cite{DBLP:journals/corr/abs-2511-12884} conducted an empirical study of agent context files, a single artifact type. \citeauthor{Seyedmoein2026}~\cite{Seyedmoein2026} studied context engineering practices across open-source repositories, analyzing the textual content of context files. Our prior study~\cite{galster2026configuring} defined a taxonomy of eight configuration mechanisms across five tools and analyzed their adoption in 2,853~repositories. That work produced findings and published supplementary scripts and CSV summaries for those repositories. This paper contributes an expanded, standalone dataset with 4,738~repositories, new data types not present in the supplementary material (148,519~AI-co-authored commits, 18,167~raw configuration files with full text content, per-artifact git metadata), and an improved pipeline to classify whether repositories belong to \emph{engineered} software projects, reducing unsure cases (i.e., repositories that could not confidently be classified) by 93\%. The raw file contents, combined with per-commit AI authorship, let researchers measure for each configuration artifact how much was written by humans versus by AI tools---a question that file presence alone cannot answer.

Several related datasets capture other aspects of AI-assisted development. The MSR 2026 Mining Challenge dataset (AIDev)~\cite{DBLP:journals/corr/abs-2602-09185} provides 932,791 agent-authored pull requests across five AI tools, focusing on code contributions rather than configurations. AgentPack~\cite{agentpack2025} collects 1.8M code edits co-authored by AI agents and humans, primarily to support training future code-editing models. \citeauthor{selfadmitted2025}~\cite{selfadmitted2025} mined explicit self-admissions of GenAI usage from commits and documentation, providing high-precision but inherently incomplete signals since many projects do not disclose AI tool usage. Repository sampling approaches such as the SEART GitHub Search (GHS) tool~\cite{DBLP:conf/msr/DabicAB21} and the criteria of \citeauthor{DBLP:journals/ese/MunaiahKCN17}~\cite{DBLP:journals/ese/MunaiahKCN17} for identifying engineered projects provide foundations for data collection, but no existing dataset applies them to AI coding tool configurations specifically. To our knowledge, no dataset provides multi-tool AI coding tool configurations with associated metadata, temporal information, and raw artifact contents.

\section{Dataset Overview}
\label{sec:dataset-overview}

Throughout this paper, the GitHub repository is the unit of analysis. A single software project may span multiple repositories (e.g., separate frontend and backend repositories). We therefore classify each repository individually as belonging to an engineered software project~\cite{DBLP:journals/ese/MunaiahKCN17}; the classification prompt extends Munaiah et al.'s original definition and is included in the pipeline archive~\cite{baltes2026pipeline} (see Section~\ref{sec:classification}).

The dataset is organized around three nested levels of repositories. At the broadest level, a \emph{sampling frame} of 40,585~GitHub repositories provides metadata for the general population of actively maintained open-source projects. Of these, 36,710 are classified as belonging to engineered software projects. At the narrowest level, our dataset contains 4,738~repositories with detected AI coding tool configurations, corresponding metadata, raw configuration file contents, and AI-co-authored commit histories. \autoref{tab:dataset-summary} lists all data files with row counts. We describe each level below.

\textbf{Sampling frame.}
All 40,585~repositories include GitHub metadata: primary programming language, star and fork counts, commit totals, contributor counts, license, creation date, and repository topics. The ten programming languages in the sampling frame are C, C\#, C++, Go, Java, JavaScript, PHP, Python, Rust, and TypeScript. Among the 4,738~repositories with detected configurations, the top five languages are TypeScript (24.5\%), Python (18.7\%), Go (13.8\%), Java (8.0\%), and C\# (7.5\%); the remaining 27.5\% is split across Rust, JavaScript, C++, PHP, and~C. This baseline allows researchers to compare tool-adopting repositories against the broader population and to control for language-specific effects.

\textbf{Classification.}
Of the 40,585~repositories, 1,159 were excluded before classification because they had become unavailable (15), lacked a \texttt{README} file (74), or had a non-English \texttt{README} (1,070). The remaining 39,426~repositories carry classification labels produced by GPT-5.2, indicating whether each repository belongs to an engineered software project. Each label includes a justification text. Of the 39,426 classified repositories, 36,710 belong to engineered software projects, 2,564 do not, and 152 are unsure cases. We discuss this process in more detail in Section~\ref{sec:classification}.

\begin{figure}[t]
    \centering
    \includegraphics[width=\columnwidth]{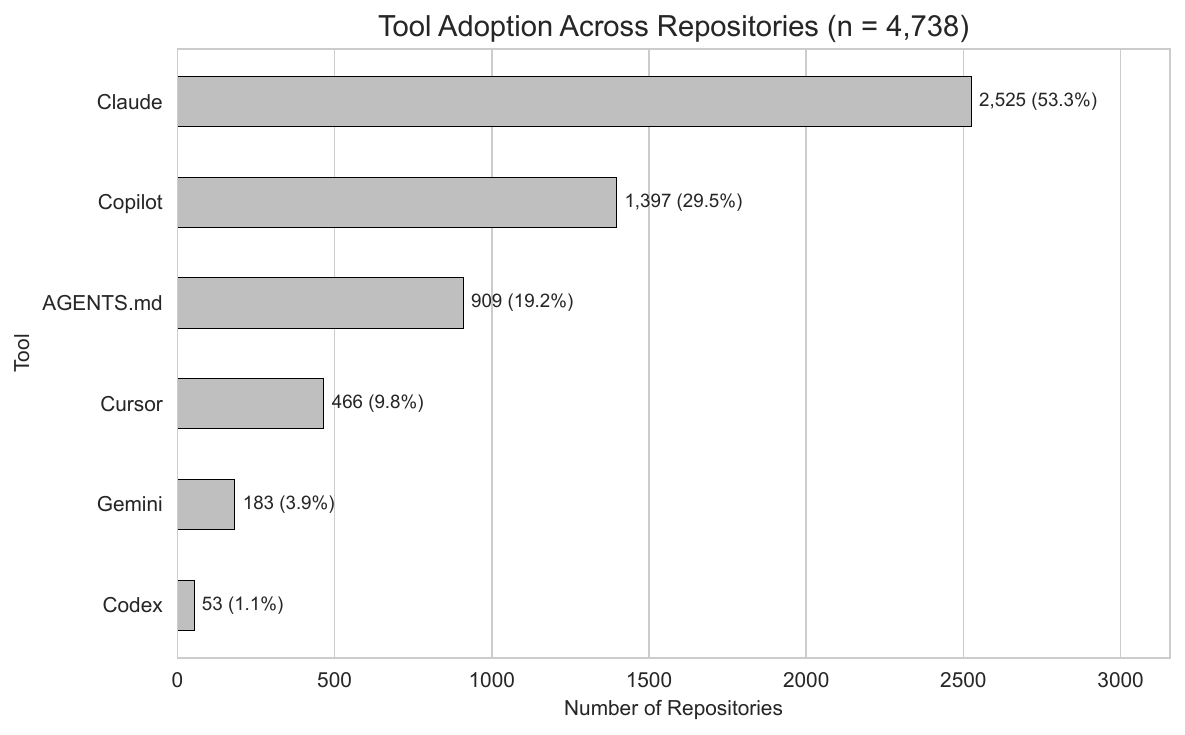}
    \caption{Number of repositories with detected configuration artifacts per agentic AI coding tool. Bars sum to more than 4,738 because a repository can configure multiple tools.}
    \Description{Bar chart showing Claude Code as the most adopted tool (2,525 repos), followed by Copilot (1,397), AGENTS.md only (909), Cursor (466), Gemini (183), and Codex (53).}
    \label{fig:tool-adoption}
\end{figure}

\begin{figure}[t]
    \centering
    \includegraphics[width=\columnwidth]{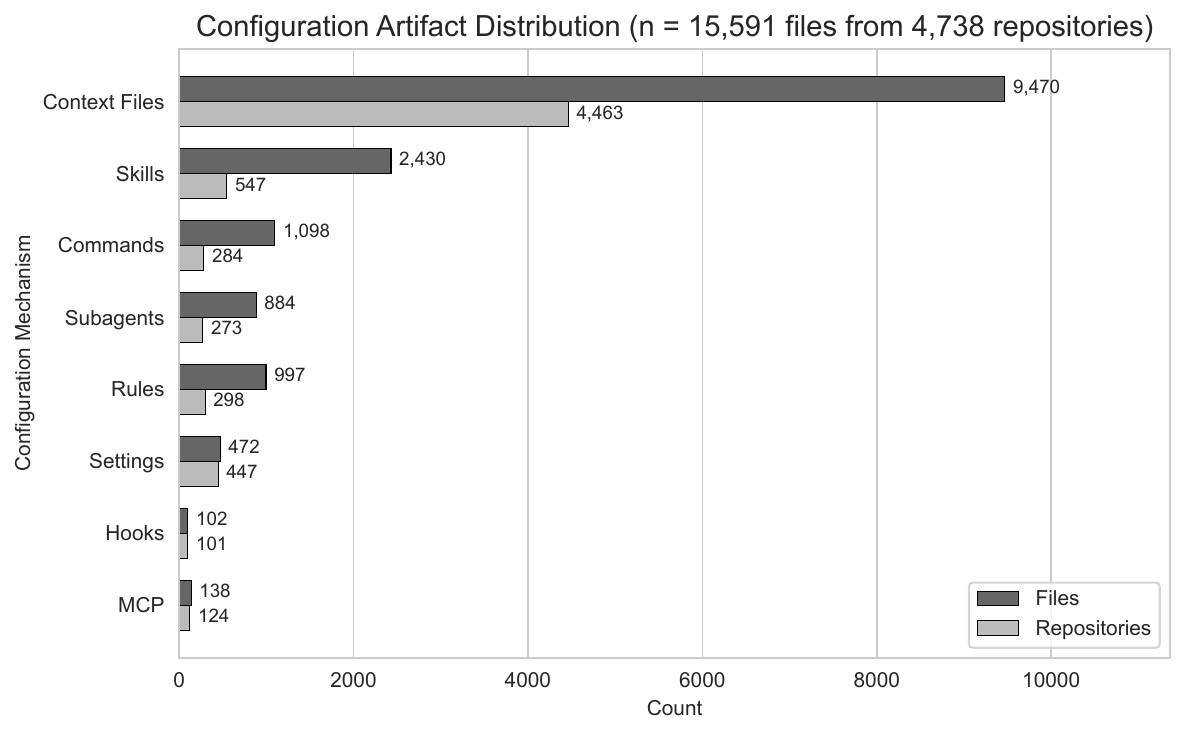}
    \caption{Number of configuration artifacts by type. For each mechanism, the chart shows the total file count and the number of repositories containing at least one artifact.}
    \Description{Bar chart showing context files as the most common artifact type with 9,470 files across 4,463 repositories, followed by skills (2,430 files, 547 repos), commands (1,098 files, 284 repos), rules (997 files, 298 repos), subagents (884 files, 273 repos), settings (472 files, 447 repos), mcp (138 files, 124 repos), and hooks (102 files, 101 repos).}
    \label{fig:artifact-distribution}
\end{figure}

\textbf{Configuration artifacts.}
Among the 36,710 repositories belonging to engineered software projects, 4,738 contain at least one agentic AI coding tool configuration artifact. The dataset covers five tools (Claude Code, GitHub Copilot, OpenAI Codex, Cursor, and Gemini) and eight configuration mechanisms that we introduced in Section~\ref{sec:introduction}: \textsc{Context Files}, \textsc{Skills}, \textsc{Subagents}, \textsc{Commands}, \textsc{Rules}, \textsc{Settings}, \textsc{Hooks}, and \textsc{MCP} configurations. \autoref{fig:tool-adoption} shows the number of repositories per tool (one repository can use more than one tool): Claude Code leads with 2,525~repositories, followed by Copilot (1,397) and Cursor (466). An additional 909~repositories contain only an \texttt{AGENTS.\allowbreak{}md} file without any tool-specific configuration, suggesting use of a tool-agnostic convention. \autoref{fig:artifact-distribution} shows the distribution of artifacts across types. \textsc{Context Files} are the most common (found in 4,463~repositories), while \textsc{Hooks} (101) and \textsc{MCP} configurations (124) remain rare.

\begin{figure}[t]
    \centering
    \includegraphics[width=\columnwidth]{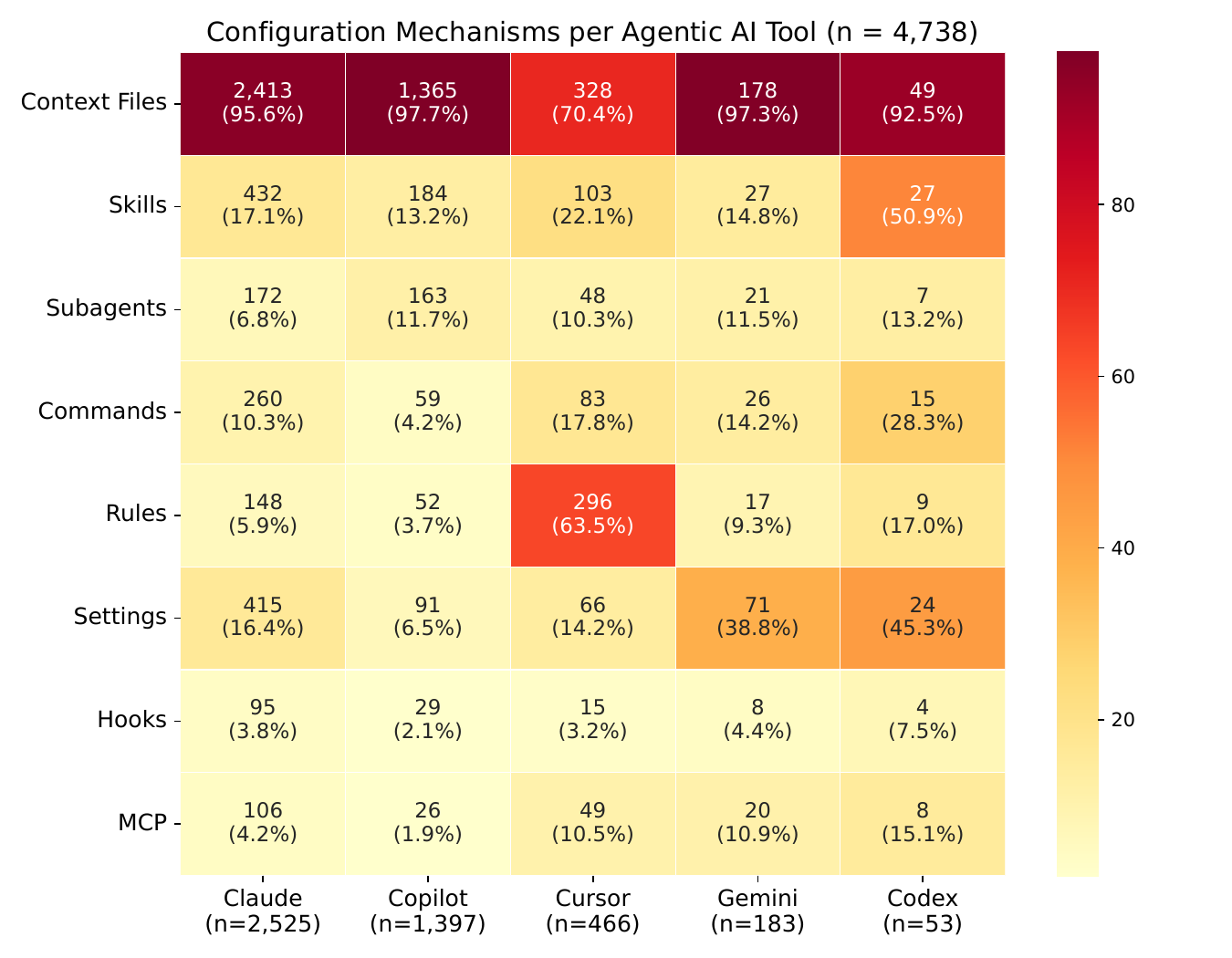}
    \caption{Configuration mechanisms per tool. Each cell shows the count and percentage of repositories using a given tool that also adopt the corresponding mechanism.}
    \Description{Heatmap showing that context files are used by nearly all repositories across all tools, while skills, subagents, and commands are concentrated in Claude Code, and rules are specific to Cursor.}
    \label{fig:config-tool-heatmap}
\end{figure}

\begin{figure*}[t]
    \centering
    \includegraphics[width=\textwidth]{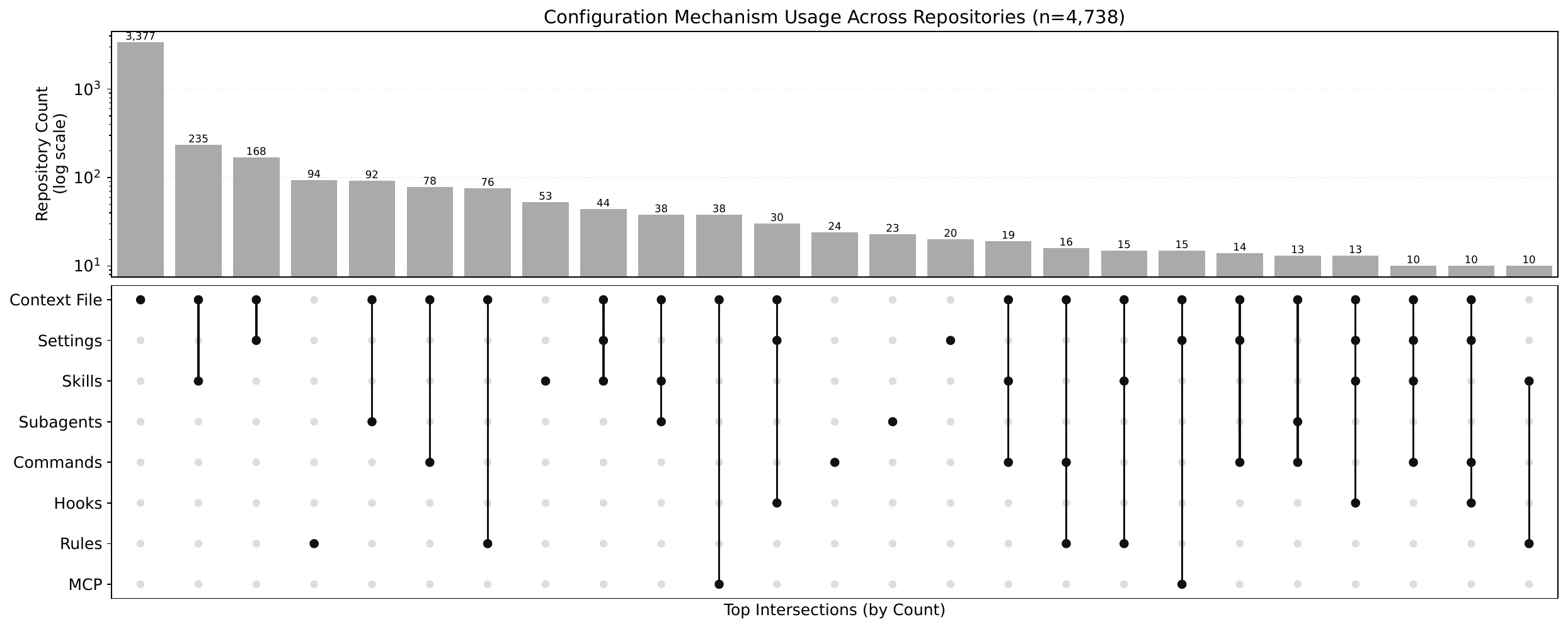}
    \caption{Co-occurrence of configuration mechanisms across repositories. The bar chart shows how many repositories use each combination; the matrix below indicates which mechanisms are included.}
    \Description{UpSet plot showing that Context File alone is the most common configuration, followed by Context File plus Settings and Context File plus Skills. Most repositories use only one or two mechanisms.}
    \label{fig:mechanism-upset}
\end{figure*}

\autoref{fig:config-tool-heatmap} breaks this down further by showing which configuration mechanisms are used with which tools. \textsc{Context Files} are near-universal across all tools. The clearest tool-specific pattern is \textsc{Rules}, adopted by 63.5\% of Cursor repositories compared to at most 17.0\% of repositories in any other tool's user base. Claude Code accounts for most \textsc{Skills}, \textsc{Subagents}, and \textsc{Commands} in absolute terms because it has the largest user base (2,525~repositories), but proportional adoption within each tool is broadly comparable: for example, Cursor and Codex users adopt \textsc{Skills} (22.1\% and 50.9\%) and \textsc{Commands} (17.8\% and 28.3\%) at rates similar to or higher than Claude Code users (17.1\% and 10.3\%). This cross-tabulation, derived directly from the dataset, illustrates the configuration practices forming around each tool. \autoref{fig:mechanism-upset} shows which combinations of configuration mechanisms co-occur: the majority of repositories use only \textsc{Context Files}, while smaller groups combine \textsc{Context Files} with \textsc{Settings}, \textsc{Skills}, or multiple advanced mechanisms.

\begin{table}[t]
\centering
\small
\caption{Dataset files and row counts. Each row in the per-mechanism CSVs (\texttt{context\_files.csv} through \texttt{hooks.csv}) is one configuration artifact, summing to 15,591 artifacts across the eight mechanisms. Large files are distributed as 7z archives.}
\label{tab:dataset-summary}
\begin{tabular}{lrl}
\toprule
\textbf{File} & \textbf{Rows} & \textbf{Description} \\
\midrule
\texttt{repos.\allowbreak{}csv} & 40,585 & Sampling frame with metadata \\
\texttt{commits.\allowbreak{}csv} & 148,519 & AI-co-authored commits \\
\texttt{context\_files.\allowbreak{}csv} & 9,470 & Context files \\
\texttt{skills.\allowbreak{}csv} & 2,430 & Skills \\
\texttt{commands.\allowbreak{}csv} & 1,098 & Commands \\
\texttt{rules.\allowbreak{}csv} & 997 & Rules \\
\texttt{subagents.\allowbreak{}csv} & 884 & Subagents \\
\texttt{settings.\allowbreak{}csv} & 472 & Settings \\
\texttt{mcp.\allowbreak{}csv} & 138 & MCP server configurations \\
\texttt{hooks.\allowbreak{}csv} & 102 & Hooks \\
\midrule
\texttt{repos\_data.\allowbreak{}7z} & -- & 18,167 raw configuration files \\
\bottomrule
\end{tabular}
\end{table}

Per-artifact CSV files record the file path, git creation date, commit count, and first and last commit SHAs for each artifact. \textsc{Context Files} include additional metadata: detected natural language, line count, whether the file references another file, and AI authorship information for the file's commits. The archive \texttt{repos\_data.\allowbreak{}7z} contains the full text of all 18,167 collected configuration files, enabling qualitative content analysis and model training. The file count (18,167) is larger than the artifact count (15,591) because some artifacts are directories: each \textsc{Skill}, \textsc{Subagent}, \textsc{Command}, or \textsc{Rule} can bundle multiple files (e.g., a Skill may pair \texttt{SKILL.\allowbreak{}md} with scripts, references, and assets). All 15,591~artifacts across all eight mechanism types were validated to be non-empty.

\textbf{AI-co-authored commits.}
For every repository with at least one configuration artifact, we collected all AI-co-authored commits (see \autoref{sec:extraction} for the detection method). The resulting \texttt{commits.\allowbreak{}csv} contains 148,519~commits from 3,392~repositories, each with a timestamp, full commit message, and the detected AI tool. For \textsc{Context Files}, the per-artifact CSV additionally records whether the file's creation commit was AI-co-authored and the total number of AI-co-authored commits that modified the file.

\section{Dataset Construction}
\label{sec:dataset-construction}

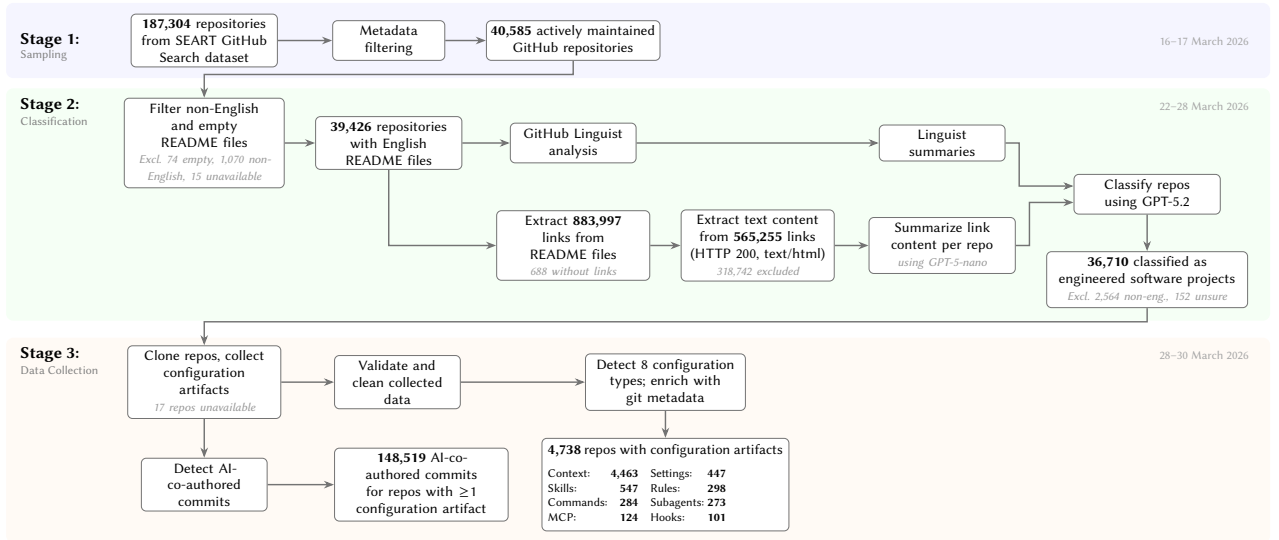
\begin{figure*}[tb]
    \centering
    \resizebox{\linewidth}{!}{
\usetikzlibrary{positioning, arrows.meta, fit, backgrounds, calc}

\newcommand{\bsub}[1]{\\ {\tiny\itshape\textcolor{black!40}{#1}}}

\begin{tikzpicture}[
    font=\sffamily\scriptsize,
    >=Stealth,
    box/.style={
        draw=black!55, rounded corners=2pt, minimum height=0.5cm,
        text width=2cm, align=center, fill=white, inner sep=2pt,
        font=\sffamily\scriptsize
    },
    arr/.style={-{Stealth[length=1.5mm]}, semithick, black!55},
    slbl/.style={font=\sffamily\footnotesize\bfseries},
    dlbl/.style={font=\sffamily\tiny, text=black!35},
]

\def\cA{2.8}   \def\cB{5.5}   \def\cC{8.2}
\def\cD{10.9}  \def\cE{13.6}  \def\cF{16.3}
\def\rI{0}       
\def\rII{-1.5}   
\def\rIII{-3.0}  
\def\rIV{-3.5}   
\def\rV{-5.0}    
\def\rVI{-6.5}   
\def\gap{0.2}    

\begin{scope}[on background layer]
    \fill[blue!4, rounded corners=3pt] (-0.1, 0.55) rectangle (18.4, -0.55);
\end{scope}
\node[slbl, anchor=west] at (0, \rI) {Stage 1:};
\node[font=\sffamily\tiny, text=black!50, anchor=west] at (0, \rI-0.22) {Sampling};
\node[dlbl, anchor=east] at (18.2, \rI) {16--17 March 2026};

\node[box] (ghs) at (\cA, \rI) {\textbf{187,304} repositories\\from SEART GitHub\\Search dataset};
\node[box, text width=1.5cm] (s1filter) at (\cB, \rI) {Metadata\\filtering};
\node[box, text width=2.4cm] (filtered) at (\cC, \rI) {\textbf{40,585} actively maintained\\GitHub repositories};

\draw[arr] (ghs) -- (s1filter);
\draw[arr] (s1filter) -- (filtered);

\begin{scope}[on background layer]
    \fill[green!4, rounded corners=3pt] (-0.1, -0.7) rectangle (18.4, -4.1);
\end{scope}
\node[slbl, anchor=west] at (0, -0.95) {Stage 2:};
\node[font=\sffamily\tiny, text=black!50, anchor=west] at (0, -1.17) {Classification};
\node[dlbl, anchor=east] at (18.2, -0.95) {22--28 March 2026};

\node[box, text width=2.2cm] (readmefilter) at (\cA, \rII) {Filter non-English\\and empty\\README files\bsub{Excl.\ 74 empty, 1,070 non-English, 15 unavailable}};
\node[box] (english) at (\cB, \rII) {\textbf{39,426} repositories\\with English\\README files};
\draw[arr] (filtered.south) -- ++(0,-\gap) -| (readmefilter.north);
\draw[arr] (readmefilter) -- (english);

\node[box, text width=1.7cm] (linguist) at (\cC, \rII) {GitHub Linguist\\analysis};
\node[box, text width=1.7cm] (lingsummary) at (\cE, \rII) {Linguist\\summaries};

\draw[arr] (english) -- (linguist);
\draw[arr] (linguist) -- (lingsummary);

\node[box, text width=2.1cm] (extractlinks) at (\cC, \rIII) {Extract \textbf{883,997}\\links from\\README files\bsub{688 without links}};
\node[box, text width=2.1cm] (extractcontent) at (\cD, \rIII) {Extract text content\\from \textbf{565,255} links\\(HTTP 200, text/html)\bsub{318,742 excluded}};
\node[box, text width=2cm] (sumlinks) at (\cE, \rIII) {Summarize link\\content per repo\bsub{using GPT-5-nano}};

\draw[arr] (english.south) |- (extractlinks.west);
\draw[arr] (extractlinks) -- (extractcontent);
\draw[arr] (extractcontent) -- (sumlinks);

\node[box] (classify) at ({\cF+0.3}, {(\rII+\rIII)/2}) {Classify repos\\using GPT-5.2};

\draw[arr] (lingsummary.east) -- ++(0.2,0) |- ($(classify.west)+(0,0.12)$);
\draw[arr] (sumlinks.east) -- ++(0.2,0) |- ($(classify.west)+(0,-0.12)$);

\node[box, text width=2.8cm] (engineered) at ({\cF+0.3}, \rIV) {\textbf{36,710} classified as\\engineered software projects\bsub{Excl.\ 2,564 non-eng., 152 unsure}};
\draw[arr] (classify) -- (engineered);

\begin{scope}[on background layer]
    \fill[orange!4, rounded corners=3pt] (-0.1, -4.35) rectangle (18.4, -7.35);
\end{scope}
\node[slbl, anchor=west] at (0, -4.6) {Stage 3:};
\node[font=\sffamily\tiny, text=black!50, anchor=west] at (0, -4.82) {Data Collection};
\node[dlbl, anchor=east] at (18.2, -4.6) {28--30 March 2026};

\node[box, text width=2.1cm] (collect) at (\cA, \rV) {Clone repos, collect\\configuration\\artifacts\bsub{17 repos unavailable}};
\node[box, text width=1.7cm, anchor=west] (validate) at ({\cB-0.8}, \rV) {Validate and\\clean collected\\data};
\node[box, text width=2.2cm] (detectenrich) at ({(\cC+\cD)/2}, \rV) {Detect 8 configuration\\types; enrich with\\git metadata};

\draw[arr] (engineered.south) -- ++(0,-\gap) -| (collect.north);
\draw[arr] (collect) -- (validate);
\draw[arr] (validate) -- (detectenrich);

\node[box, text width=1.7cm] (aicommits) at (\cA, \rVI) {Detect AI-co-authored\\commits};
\node[box, text width=2.4cm, anchor=west] (airesult) at ({\cB-0.8}, \rVI) {\textbf{148,519} AI-co-authored commits\\for repos with $\geq$1 configuration artifact};

\node[draw=black!55, rounded corners=2pt, fill=white, inner sep=2.5pt,
      font=\sffamily\tiny, align=left] (output) at ({(\cC+\cD)/2}, \rVI) {%
    \textbf{\scriptsize 4,738} {\scriptsize repos with configuration artifacts}\\[3pt]
    \begin{tabular}{@{}l@{\;}r@{\quad}l@{\;}r@{}}
    Context: & \textbf{4,463} & Settings: & \textbf{447} \\
    Skills: & \textbf{547} & Rules: & \textbf{298} \\
    Commands: & \textbf{284} & Subagents: & \textbf{273} \\
    MCP: & \textbf{124} & Hooks: & \textbf{101} \\
    \end{tabular}%
};

\draw[arr] (collect.south) -- (aicommits.north);
\draw[arr] (aicommits) -- (airesult);
\draw[arr] (detectenrich.south) -- (output.north);

\end{tikzpicture}}
    \caption{Three-stage data collection pipeline: repository sampling from the SEART GitHub Search dataset, LLM-based classification of engineered software projects, and extraction of AI coding tool configuration artifacts and AI-co-authored commits.}
    \Description{Diagram showing the three-stage data collection pipeline: sampling 187,304 repositories from SEART GHS, using GPT-5.2 to identify 36,710 of the 39,426 classified repositories as belonging to engineered software projects, and detecting AI configuration artifacts in 4,738 repositories along with 148,519 AI-co-authored commits.}
    \label{fig:data-collection}
\end{figure*}

\looseness=-1
The dataset was constructed through a three-stage pipeline: repository sampling, LLM-based classification, and configuration artifact extraction. \autoref{fig:data-collection} illustrates the pipeline with counts at each stage.

\subsection{Repository Sampling}
\label{sec:sampling}

We used the GHS\footnote{\url{https://seart-ghs.si.usi.ch/}} dataset~\cite{DBLP:conf/msr/DabicAB21} as a starting point. GHS provides a regularly updated snapshot of GitHub repository metadata collected via the GitHub API. At the time of our query, it contained over one million repositories with metadata covering repository information, activity metrics, and social engagement indicators.

Our baseline query was designed to avoid common pitfalls in GitHub mining~\cite{DBLP:conf/msr/KalliamvakouGBSGD14}. A repository in our sample had to (1)~have a license, (2)~not be a fork, (3)~have at least two contributors, (4)~have at least one pull request, (5)~be at least 18 months old, and (6)~have had a commit within the past six months. GHS additionally includes only repositories with at least ten stars. Requiring a license enables downstream open-source filtering. Excluding forks avoids duplicated codebases. The contributor and pull request thresholds exclude toy and solo projects. The age and recency requirements filter both immature and stale repositories.

The query, executed on 2026-03-16, returned 187,304~repositories. We then applied metadata filtering stages (\autoref{tab:filtering-table}): (1) removing archived, disabled, or locked repositories; (2) retaining only OSI-approved software licenses per the SPDX License List; (3) keeping only the ten most common programming languages; (4) excluding educational and resource-aggregation repositories, identified by regular-expression patterns over the repository name and the repository's GitHub topics (as recorded by GHS) that match keywords such as \emph{tutorial}, \emph{course}, \emph{awesome-*}, and \emph{list}; (5) applying median-based lower thresholds on commits (${\geq}352$) and watchers (${\geq}6$)~\cite{Seyedmoein2026}; (6) deduplicating redirected repositories. The full regular expressions are in the pipeline archive's YAML configuration~\cite{baltes2026pipeline}. After filtering, 40,585~repositories remained.

\begin{table}[t]
\centering
\small
\caption{Metadata filtering stages and their impact on dataset size, applied to the baseline GHS query result of 187,304 repositories.}
\label{tab:filtering-table}
\begin{tabularx}{\columnwidth}{Xrr}
\toprule
\textbf{Filtering Step} & \textbf{\#Excluded} & \textbf{\#Remaining} \\
\midrule
Initial dataset                                                & --     & 187,304 \\
(1) Excluded archived, disabled, or locked repositories        & 2,241  & 185,063 \\
(2) Excluded repositories without OSI-approved licenses        & 24,646 & 160,417 \\
(3) Excluded repositories outside the top-10 languages         & 31,252 & 129,165 \\
(4) Excluded educational and resource repositories             & 1,994  & 127,171 \\
(5) Excluded repositories with $<352$ commits or $<6$ watchers & 86,331 & 40,840 \\
(6) Excluded duplicates of redirected repositories             & 255    & \textbf{40,585} \\
\bottomrule
\end{tabularx}
\end{table}

\subsection{Repository Classification}
\label{sec:classification}

We classified the sampled repositories to distinguish those belonging to engineered software projects from non-software or toy repositories. After cloning each repository, we excluded 15 that were unavailable, 74 with missing or empty \texttt{README} files, and 1,070 with non-English \texttt{README}s (detected using \texttt{lingua-language-detector}), leaving 39,426~repositories.

For each repository, we assembled three types of inputs for the classification: (1)~the \texttt{README} content, (2)~a summary of the file structure produced by GitHub Linguist, and (3)~summaries of external links referenced in the \texttt{README}.

For the file structure, we ran GitHub Linguist on each repository and summarized the output by retaining the top seven languages per repository (ranked by byte count; covering the 75th percentile of the language-per-repository distribution) and up to three representative file paths per directory level, capped at depth eight (where 75\% of all files across our sample reside). This keeps the representation informative without exceeding the context window of the model used for classification (see below).

For external links, we extracted all URLs from the \texttt{README} files (883,997~links across 39,426~repositories). We retained only those returning HTTP~200 with \texttt{text/\allowbreak{}html} content type (565,255~links) and summarized the first 24 per repository (the 75th percentile of links per repository) using GPT-5-nano. \citeauthor{DBLP:journals/ese/PranaTTAL19}~\cite{DBLP:journals/ese/PranaTTAL19} found that links in \texttt{README} files cluster in the ``What'' and ``How'' sections that typically appear near the top of the file, so earlier links tend to be more informative about the repository's purpose.

We used GPT-5.2 via the OpenAI Batch API to classify each repository based on these three inputs. The classification prompt extends \citeauthor{DBLP:journals/ese/MunaiahKCN17}'s~\cite{DBLP:journals/ese/MunaiahKCN17} definition of engineered software projects with positive indicators (e.g., stated purpose, installation instructions, CI/CD) and explicit exclusion criteria for categories such as tutorials, scaffolding outputs, and code-snippet collections. Two authors spot-checked the assigned labels on a random sample. We piloted both GPT-5.2 and GPT-5-nano; GPT-5.2 showed better adherence to the classification criteria. The exact classification prompt is included in the pipeline archive~\cite{baltes2026pipeline}.

The classification labeled 36,710~repositories as belonging to engineered software projects, 2,564 as not belonging to such projects, and 152 as unsure. Compared to our prior study~\cite{galster2026configuring}, which classified repositories based on \texttt{README} content alone, this improved pipeline reduced unsure cases from 2,204 to 152. The reduction is attributable to the additional context from file structure summaries and external link summaries, which helped resolve borderline cases.

\subsection{Configuration Artifact Extraction}
\label{sec:extraction}

We selected five agentic AI coding tools based on the 2025 Stack Overflow Developer Survey~\cite{stackoverflow-survey-2025}: Claude Code, GitHub Copilot, OpenAI Codex, Cursor, and Gemini. The survey asked developers which AI agent tools they use; we selected the four most popular tools and substituted Codex for ChatGPT, as Codex is the agentic coding tool by the same vendor. We included Cursor given its prominence in recent software engineering research~\cite{he2026speed, jiang2026beyond}.

For each tool, we documented the repository-level files and directories that indicate its presence and developed detection heuristics based on this documentation~\cite{galster2026configuring}. We detect the eight configuration mechanisms introduced in Section~\ref{sec:introduction}; example artifacts include \texttt{CLAUDE.\allowbreak{}md}, \texttt{AGENTS.\allowbreak{}md}, and \texttt{.\allowbreak{}cursorrules} for \textsc{Context Files}.
For GitHub Copilot, Cursor, and Gemini, the detected files apply to both their conversational and agentic interfaces.

We cloned the 36,710 repositories belonging to engineered software projects and applied these heuristics to extract configuration artifacts. After validation and cleaning (removing invalid file paths, recalculating detection flags, and excluding repositories where no valid artifacts remained), 4,738 repositories contained at least one artifact. We enriched each artifact with git metadata: creation date, total commit count, and first and last commit hashes. For \textsc{Context Files}, we additionally detected the natural language, identified files that reference other files, and attributed AI authorship at the commit level.

\looseness=-1
To assess the extent of active AI tool use in these repositories, we scanned each repository's full commit history for AI-co-authored commits. Detection uses regex-based heuristics across three scopes, applied to every commit reachable from any branch. First, \emph{author identity} patterns match tool-specific bot accounts in the author and committer name and email fields (e.g., \texttt{copilot-swe-agent[bot]}, \texttt{claude[bot]}). Second, \emph{git trailer} patterns match structured attribution lines such as \texttt{Co-authored-by:} and \texttt{Generated-by:}, extracted by walking the commit body in reverse to isolate the trailer block from the message body. Third, \emph{message body} patterns match attribution phrases and tool-specific branch prefixes in merge commits. The patterns cover five tools (Claude, Copilot, Cursor, Codex, and Gemini) and are documented with unit tests in the pipeline archive~\cite{baltes2026pipeline}. This produced 148,519~AI-co-authored commits across 3,392 of the 4,738~configured repositories.

\subsection{Dataset Validation}
\label{sec:validation}

Four independent signals from the pipeline's outputs indicate that the detected artifacts correspond to real AI-coding configurations. First, the sampling frame already excludes non-software populations. Repositories must have one of ten mainstream programming languages (C, C\#, C++, Go, Java, JavaScript, PHP, Python, Rust, TypeScript) as their primary language, and the GPT-5.2 classification further filters to repositories belonging to engineered software projects (see Section~\ref{sec:classification}). Files such as \texttt{AGENTS.\allowbreak{}md} also appear outside coding-agent contexts (e.g., in agent-framework experiments or prompt libraries), but such uses concentrate in repositories that these filters exclude. Second, 71.6\% of the 4,738~configured repositories (3,392) contain at least one AI-co-authored commit, a lower bound because commit detection requires explicit attribution signals that not every tool or workflow leaves behind. Third, each commit is checked against three independent signals: the author identity, git trailers, and patterns in the commit message body. Author-identity matches correspond to verified bot accounts such as \texttt{copilot-swe-agent[bot]}. Fourth, the prior study~\cite{galster2026configuring} manually inspected every short configuration file ($\leq$10 lines) in its own 2,853-repo sample and found that 87.8\% (396 of 451) contained substantive configuration content. The per-mechanism breakdown is preserved in the pipeline archive~\cite{baltes2026pipeline}. Per-repository justification texts are shipped with the dataset so users can re-audit any label.

\section{Use Cases}
\label{sec:use-cases}

The dataset supports a range of empirical research questions, each illustrated below with the data fields that enable it.

\begin{figure*}[t]
    \centering
    \includegraphics[width=0.8\textwidth]{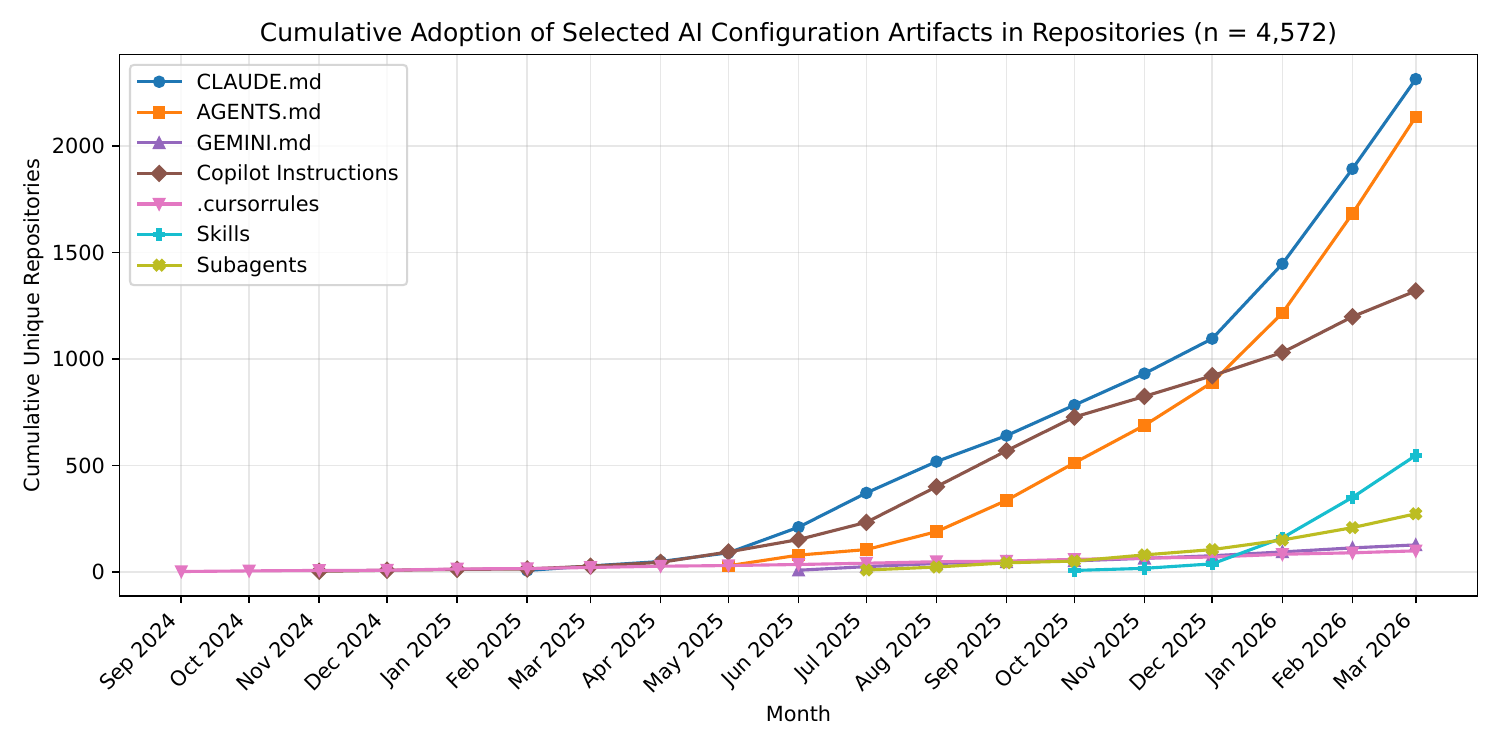}
    \caption{Cumulative adoption of selected AI configuration artifacts over time. Each curve shows the cumulative number of repositories in which the artifact type was first observed.}
    \Description{Line chart showing cumulative adoption curves for CLAUDE.md, AGENTS.md, GEMINI.md, Copilot instructions, .cursorrules, Skills, and Subagents from 2023 to 2026, with CLAUDE.md and AGENTS.md showing the steepest growth.}
    \label{fig:cumulative-adoption}
\end{figure*}

\textbf{Tool adoption and evolution.}
Each configuration artifact in the dataset carries a \texttt{created\_at} timestamp and commit count, and each AI-co-authored commit has a timestamp. Researchers can use these fields to study when repositories adopted specific tools, in what order they added configuration mechanisms, and whether tool adoption correlates with changes in commit activity. \autoref{fig:cumulative-adoption} shows the cumulative number of repositories adopting each artifact type, illustrating the steep growth of \texttt{CLAUDE.\allowbreak{}md} and \texttt{AGENTS.\allowbreak{}md} from late 2024 and the later adoption of \textsc{Skills} and \textsc{Subagents}. 

\textbf{Configuration practices.}
\looseness=-1
The raw configuration files in \texttt{repos\_\allowbreak{}data.\allowbreak{}7z} allow qualitative and quantitative content analysis. What instructions do developers give AI agents? Do configuration patterns differ by programming language or repository size? How do configuration files for different tools compare in length, structure, and content? Researchers can also compare how the eight configuration mechanisms are combined within repositories: our prior study~\cite{galster2026configuring} found that most repositories use only \textsc{Context Files}, while advanced mechanisms such as \textsc{Skills} and \textsc{Subagents} remain rare.

\begin{figure}[t]
    \centering
    \includegraphics[width=\columnwidth]{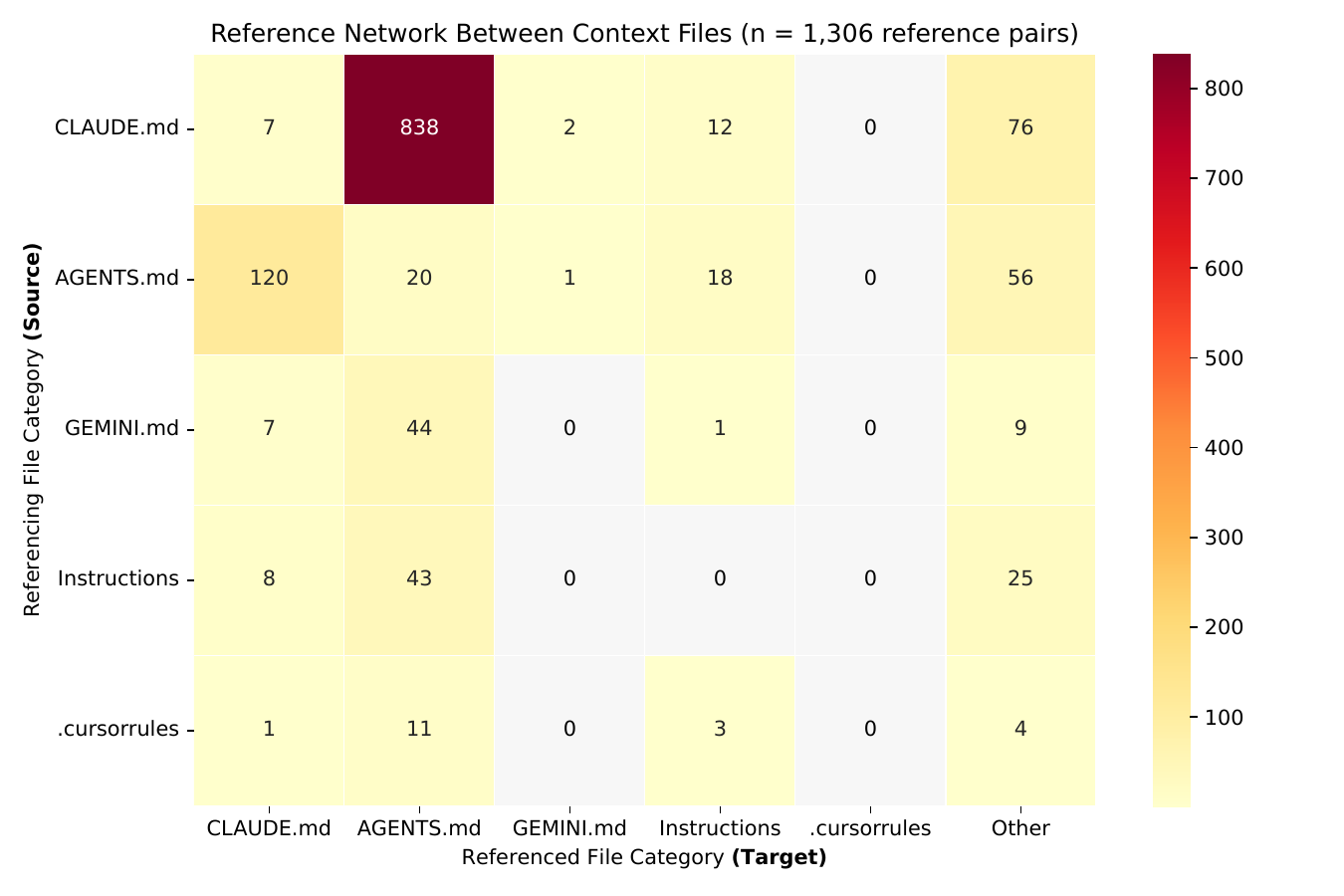}
    \caption{Reference network between context file types. Each cell shows how many files of the source type (row) reference a file of the target type (column).}
    \Description{Heatmap showing that AGENTS.md is the dominant reference target, receiving incoming references from CLAUDE.md, Copilot instructions, .cursorrules, and GEMINI.md context files.}
    \label{fig:reference-network}
\end{figure}

The \textsc{Context File} metadata records whether a file is a reference to another file (e.g., a \texttt{CLAUDE.\allowbreak{}md} that simply points to a shared \texttt{AGENTS.\allowbreak{}md}). \autoref{fig:reference-network} shows the resulting reference network: \texttt{AGENTS.\allowbreak{}md} is the dominant reference target, receiving incoming references from tool-specific context files of all five tools. This pattern indicates that \texttt{AGENTS.\allowbreak{}md} is adopted across tools as a tool-agnostic standard that tool-specific files reference.

\begin{figure*}[t]
    \centering
    \includegraphics[width=0.8\textwidth]{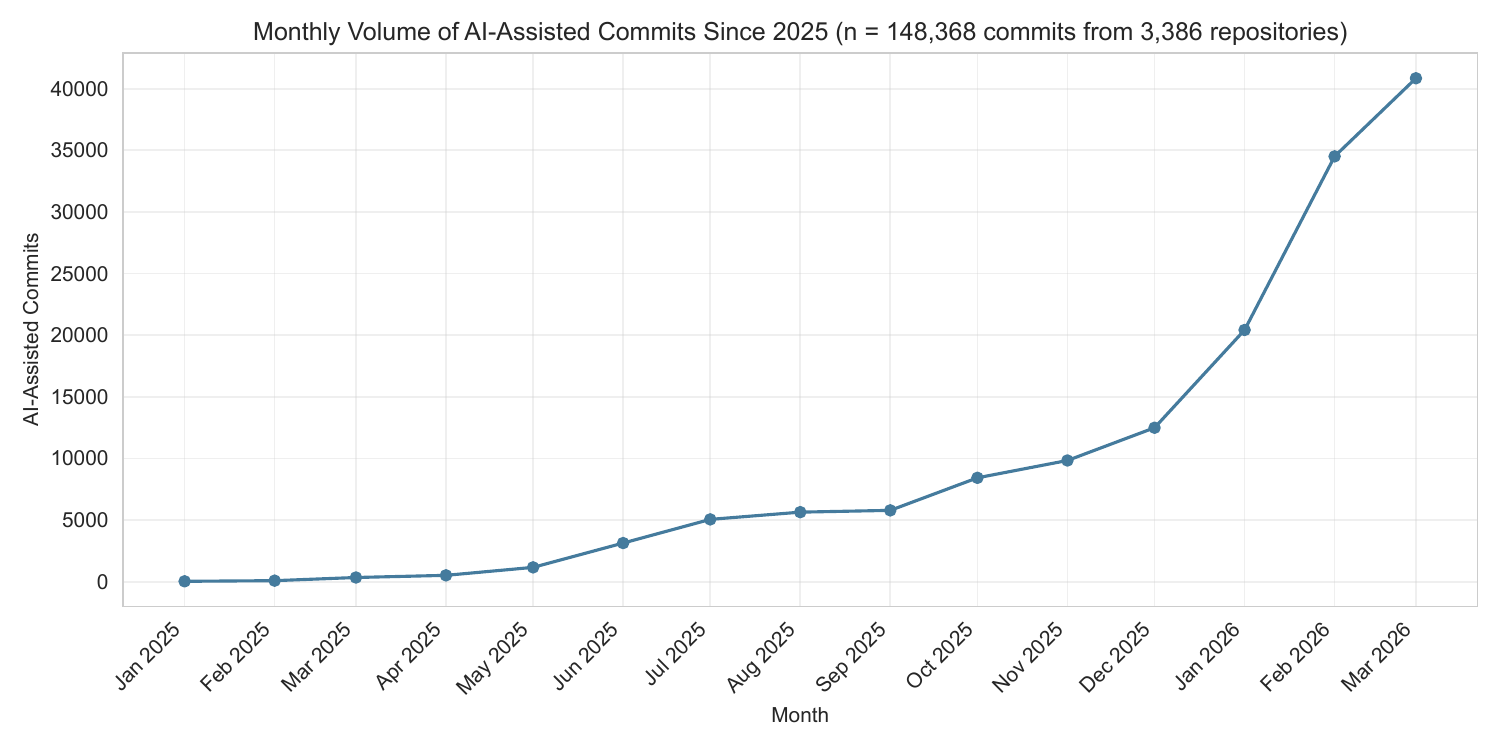}
    \caption{Monthly volume of AI-co-authored commits across the 4,738 repositories with detected configuration artifacts.}
    \Description{Line chart showing monthly AI-co-authored commit volume from 2024 to early 2026, with steep growth through 2025 into 2026.}
    \label{fig:monthly-ai-commits}
\end{figure*}

\textbf{AI-co-authored commits.}
The \texttt{commits.\allowbreak{}csv} file enables analysis of AI-co-authored commits: their frequency, timing, commit message style, and distribution across repositories. \autoref{fig:monthly-ai-commits} shows the corresponding growth in AI-co-authored commits, with monthly volumes rising from near zero in early 2025 to over 40,000 by March 2026. The rise coincides with the availability of agentic features in Claude Code and GitHub Copilot, and suggests that configuration artifact adoption and active AI tool usage are closely linked.

\looseness=-1
Each commit record includes the full commit message and the detected AI tool, allowing researchers to compare the writing style and scope of AI-generated commits with human-authored ones. Researchers can also measure what fraction of a repository's total commits are AI-co-authored, and how this fraction changes over time.

\textbf{Configuration authorship evolution.}
Researchers can join \texttt{repos\_data.\allowbreak{}7z} (raw file contents at HEAD) with \texttt{commits.\allowbreak{}csv} and each configuration artifact's first and last commit SHAs to trace whether a \texttt{CLAUDE.\allowbreak{}md}, \texttt{AGENTS.\allowbreak{}md}, or other artifact was originally authored by a human or by an AI tool and how it evolved afterwards. The \texttt{context\_files.\allowbreak{}csv} metadata already surfaces the first-commit author and a per-file AI-commit count; \texttt{git blame} against the stored files yields line-level attribution. This supports questions such as: What fraction of typical \texttt{CLAUDE.\allowbreak{}md} content is human-written versus AI-written? Do AI-authored additions focus on mechanical details (build commands, test runners) while humans write architectural guidance? Which configuration mechanisms are most often initialized by humans and later extended by AI tools, and which ones are AI-initiated from the start?

\textbf{Repository characteristics.}
Because \texttt{repos.\allowbreak{}csv} covers both tool-adopting and non-adopting repositories with full GitHub metadata, researchers can study what distinguishes repositories that adopt AI coding tools, using metrics such as repository size, contributor count, language, and activity level. The classification justifications in \texttt{repos.\allowbreak{}csv} can also serve as training data for automated repository classification approaches.

\textbf{Governance and safety.}
The raw configuration files encode how developers constrain AI agent behavior: which files agents may modify, which actions require human approval, and what security or coding practices agents must follow. Researchers can study how governance and safety constraints are expressed across repositories, identify under-specified areas, and detect risks such as ambiguity, over-delegation, or missing safeguards. Repositories that configure multiple tools simultaneously allow studying whether instructions across artifacts are consistent or contradictory.

\textbf{Model training and evaluation.}
The 18,167 configuration files can serve as a corpus for training or fine-tuning language models on developer-written instructions for AI agents. The commit-level AI authorship metadata lets researchers filter for human-authored configurations when building training sets. The per-artifact metadata (creation date, commit count, language) supports stratified sampling across these dimensions.

\section{Reproducibility and Availability}
\label{sec:reproducibility}

The dataset is archived on Zenodo under CC~BY~4.0~\cite{baltes2026dataset}. It includes all CSV files listed in \autoref{tab:dataset-summary} and the \texttt{repos\_data.\allowbreak{}7z} archive containing the raw text of all 18,167 collected configuration files. Large files (\texttt{repos.\allowbreak{}csv} and \texttt{commits.\allowbreak{}csv}) are additionally distributed as 7z archives for faster download. Users can load the data directly with standard tools and join across files:
\begin{lstlisting}
import pandas as pd
repos = pd.read_csv("repos.csv")
configured = repos[repos["scanned_at"].notna()]
claude_repos = configured[configured["claude"] == True]
commits = pd.read_csv("commits.csv")
claude_commits = commits[commits["ai_tool"].str.contains("Claude")]
\end{lstlisting}

The complete construction pipeline (all Python scripts and YAML configuration files across the three stages shown in Figure~\ref{fig:data-collection}) is archived separately on Zenodo~\cite{baltes2026pipeline}. The pipeline is organized into four directories matching the methodology sections. 
Each stage reads from the previous stage's output and produces self-contained results. The sampling stage is driven by a declarative YAML configuration file that specifies all filter parameters, making queries reproducible given the same GHS snapshot. The classification stage depends on the OpenAI API (GPT-5.2 and GPT-5-nano), introducing potential non-determinism across runs; we include the classification justifications in the dataset so that users can inspect and verify individual labels. The artifact extraction stage uses deterministic file-matching heuristics documented in a versioned specification, with unit tests for the AI commit detection logic.

An interactive website\footnote{\url{https://se-uhd.de/ai-config-dataset/}} provides a searchable interface for browsing repositories, viewing configuration artifacts, exploring AI-co-authored commits, and consulting usage examples without downloading the full dataset.

\section{Limitations}
\label{sec:limitations}

As with any large-scale mined dataset, this dataset has limitations stemming from sampling decisions, automated classification, and heuristic-based extraction.

\looseness=-1
The dataset covers only public GitHub repositories and does not include proprietary codebases, where agentic AI coding tools may be adopted at a larger scale and under different organizational constraints. AI tool adoption on other platforms (GitLab, Bitbucket) may follow different patterns; extending the pipeline to other platforms would require adapting the repository sampling and cloning stages.

LLM-based classification of engineered projects introduces non-determinism: re-running GPT-5.2 on the same input may produce slightly different labels. We include justification texts in the dataset, and two authors spot-checked labels on a random sample. The 152~unsure repositories are included in \texttt{repos.\allowbreak{}csv} with their labels, so users can apply their own inclusion criteria.

\looseness=-1
Tool detection relies on file and directory naming conventions documented by each tool's vendor. Repositories that use non-standard locations or that removed configuration files before our March 2026 snapshot are missed. The detection heuristics are versioned in the pipeline archive and can be extended as tools evolve.

AI-co-authored commit detection depends on identifiable markers in author fields, git trailers, and commit messages. Tools or workflows that do not leave such traces produce false negatives, meaning the true volume of AI-assisted contributions is likely underestimated. Some tools do not expose consistent attribution signals, which points to the need for more systematic approaches to identify AI-authored commits regardless of whether explicit attribution is present. The detection patterns are documented and extensible.
The dataset reflects a single point-in-time snapshot (March 2026). We designed the pipeline for periodic re-execution with updated data.

\section{Conclusion}
\label{sec:conclusion}

We present a curated dataset of agentic AI coding tool configurations from 4,738~open-source GitHub repositories, covering five tools, eight configuration mechanisms, 15,591~configuration artifacts with their full content, and 148,519~AI-co-authored commits. The dataset and the construction pipeline are publicly available on Zenodo. An interactive website provides detailed usage examples and an alternative entry point for browsing and exploring the dataset without downloading it.

We intend to continuously update the dataset as new tools and configuration mechanisms are introduced, expanding the detection heuristics accordingly. Future versions will also improve the classification pipeline by incorporating quantitative repository metadata (stars, forks, activity levels) and the GitHub repository description alongside the \texttt{README} content, which should further reduce the number of unsure cases. Beyond maintenance, future work can link configuration practices to downstream outcomes such as code quality and review efficiency.

\balance
\bibliographystyle{ACM-Reference-Format}
\bibliography{literature}

\end{document}